\documentclass[conference]{IEEEtran}

 \usepackage{amsmath,amssymb,amsfonts}
 \usepackage{textcomp}
\usepackage{xspace}
\usepackage{balance}
\usepackage{listings}
\usepackage[normalem]{ulem} 
\usepackage{array, booktabs}
\usepackage{blindtext}
\usepackage{ifthen}

\usepackage{csvsimple,supertabular}
\usepackage[most]{tcolorbox}
\usepackage[caption=false]{subfig}
\usepackage{paralist} 
\usepackage{parskip}
\usepackage{tabularx}

\usepackage{fontawesome}      
\usepackage{adjustbox} 		
\usepackage{pifont} 		

\newcolumntype{R}[2]{
    >{\adjustbox{angle=#1,lap=\width-(#2),margin*=0.4em 0em 0em 0em}\bgroup}
    l
    <{\egroup}
}

\let\oldFootnote\footnote
\newcommand\nextToken\relax
\renewcommand\footnote[1]{%
    \oldFootnote{#1}\futurelet\nextToken\isFootnote}
\newcommand\isFootnote{%
    \ifx\footnote\nextToken\textsuperscript{,}\fi}

\newcommand{\id}[1]{$-$Id: scgPaper.tex 32478 2010-04-29 09:11:32Z oscar $-$}

\newcommand{\ie}{\emph{i.e.},\xspace}
\newcommand{\eg}{\emph{e.g.},\xspace}
\newcommand{\etal}{\emph{et al.}\xspace}

\newcommand{\smallfootnote}[1]{\footnote{\fontsize{6}{6}\selectfont #1}}

\newcolumntype{L}[1]{>{\raggedright\arraybackslash}m{#1}} 

\def\BibTeX{{\rm B\kern-.05em{\sc i\kern-.025em b}\kern-.08em
	T\kern-.1667em\lower.7ex\hbox{E}\kern-.125emX}}
	
\newboolean{showedits}
\setboolean{showedits}{true} 
\ifthenelse{\boolean{showedits}}
{
	\newcommand{\del}[1]{\textcolor{red}{\sout{#1}}} 
	\newcommand{\nbe}[3]{
		{\colorbox{#3}{\bfseries\sffamily\scriptsize\textcolor{white}{#1}}}
		{\textcolor{#3}{\sf\small$\blacktriangleright$\textit{#2}$\blacktriangleleft$}}}
}{
	\newcommand{\del}[1]{} 
	
	\newcommand{\nbe}[3]{}
}

\ifthenelse{\boolean{showedits}}
{
	\newtcolorbox{inserted}{
		title=Inserted text:,
		colframe=blue,colback=blue!5!white,
		breakable,
		leftrule=0mm, 
		bottomrule=0mm,
		rightrule=0mm,
		toprule=0mm,
		arc=0mm, outer arc=0mm,
		oversize
	}
	\newtcolorbox{deleted}{
		title=Deleted text:,
		colframe=red,colback=red!5!white,
		breakable,
		leftrule=0mm, 
		bottomrule=0mm,
		rightrule=0mm,
		toprule=0mm,
		arc=0mm, outer arc=0mm,
		oversize
	}
	\newtcolorbox{refactored}{
		title=Rewritten text:,
		colframe=blue,colback=red!5!white,
		breakable,
		leftrule=0mm, 
		bottomrule=0mm,
		rightrule=0mm,
		toprule=0mm,
		arc=0mm, outer arc=0mm,
		oversize
	}
}{

}

\newboolean{showcomments}
\setboolean{showcomments}{true}
\ifthenelse{\boolean{showcomments}}
{\newcommand{\nbc}[3]{
		{\colorbox{#3}{\bfseries\sffamily\scriptsize\textcolor{white}{#1}}}
		{\textcolor{#3}{\sf\small$\blacktriangleright$\textit{#2}$\blacktriangleleft$}}}
	}
{\newcommand{\nbc}[3]{}
	}

\definecolor{source}{gray}{0.9}
\newboolean{isblinded}
\setboolean{isblinded}{true}
\ifthenelse{\boolean{isblinded}}
{\newcommand\blind[1]{BLINDED\xspace}}
{\newcommand\blind[1]{#1\xspace}}

\newcommand{\seclabel}[1]{\label{sec:#1}}

\newcommand{\figlabel}[1]{\label{fig:#1}}
\newcommand{\figref}[1]{\autoref{fig:#1}}
\newcommand{\tablabel}[1]{\label{tab:#1}}
\newcommand{\tabref}[1]{\autoref{tab:#1}}

\usepackage[pdftex,colorlinks=true,pdfstartview=FitV,linkcolor=black,citecolor=black,urlcolor=black,bookmarks=false]{hyperref}

\newcommand{\crule}[1]{\emph{``#1''}\xspace}

\begin{document}

\title{Do Comments follow Commenting Conventions? \\ A Case Study in Java and Python}
\author{
\IEEEauthorblockN{
Pooja Rani\IEEEauthorrefmark{1},
Suada Abukar\IEEEauthorrefmark{1},
Nataliia Stulova\IEEEauthorrefmark{1},
Alexandre Bergel\IEEEauthorrefmark{2},
Oscar Nierstrasz\IEEEauthorrefmark{1}}
\IEEEauthorblockA{
    \IEEEauthorrefmark{1}Software Composition Group, University of Bern, Bern, Switzerland\\
    \IEEEauthorrefmark{2}Department of Computer Science (DCC), University of Chile, Santiago, Chile
\\\href{http://scg.unibe.ch/staff/}{\faGlobe \hspace{0.1cm}scg.unibe.ch/staff}}
}

\maketitle
\begin{abstract}
Assessing code comment quality is known to be a difficult problem. 
A number of coding style guidelines have been created with the aim to encourage writing of informative, readable, and consistent comments.
However, it is not clear from the research to date which specific aspects of comments the guidelines cover (\eg syntax, content, structure).
Furthermore, the extent to which developers follow these guidelines while writing code comments is unknown.
    
We analyze various style guidelines in Java and Python and uncover that the majority of them address more the content aspect of the comments rather than syntax or formatting.
However, when considering the different types of information developers embed in comments and the concerns they raise on various online platforms about the commenting practices, existing comment conventions are not yet specified clearly enough, nor do they adequately cover important concerns.
We find that developers of both languages follow the writing style and content-related comment conventions more often than syntax and structure types of conventions.
Our results highlight the mismatch between developer commenting practices and style guidelines, and provide several focal points for the design and improvement of comment quality checking tools.
\end{abstract}

\begin{IEEEkeywords}
Comment analysis, Software documentation, Coding Style Guidelines, Coding Standards
\end{IEEEkeywords}

\section{Introduction}
Developers use several kinds of software documentation, including design documents, wikis, and code comments, to understand and maintain programs.
Studies show that developers trust code comments more than other forms of documentation~\cite{Maal14a}.
As code comments are usually written in a semi-structured manner using natural language sentences, and they are not checked by the compiler, developers have the freedom to write comments in various ways~\cite{Alla14a,Pasc17a,Zhan18a}.

To encourage developers to write consistent, readable, and informative code comments, programming language communities and several large organizations, such as Google and Apache,
 provide coding style guidelines that also suggest comment-related conventions~\cite{Orac21a,Pyth21a,Goog21a}.
These conventions cover various aspects of comments, such as syntactic, stylistic, or content-related aspects.
For instance, \crule{Use 3rd person (descriptive), not 2nd person (prescriptive)} is an example of a stylistic comment convention for Java documentation comments~\cite{Orac21a}.
However, to what extent these aspects are covered within different style guidelines and languages is not known.
Therefore, we formulate the question: 
\emph{\textbf{RQ$_1$}: Which type of comment conventions are suggested by various style guidelines?}

As high-quality comments support developers in understanding and maintaining their programs, it is essential to ensure the adherence of their comments to the style guidelines to evaluate the overall comment quality.
Rani~\etal have investigated class comments of Smalltalk and their adherence to the commenting conventions provided by a default template~\cite{Rani21b}.
They found that Smalltalk developers follow writing style and content-related comment conventions more than 50\% of the time, but they use inconsistent structure and formatting of comment content.
As Java and Python are among the most popular languages in use, several research works have focused on studying comments in Java and Python~(\cite{Pasc17a,Zhan18a}), some especially focusing on class comments~\cite{Rani21d}.
However, it remains largely unknown whether Java and Python developers adhere to the commenting conventions suggested by the style guidelines or not.
To obtain this understanding, we formulate another research question:
\emph{\textbf{RQ$_2$}: To what extent do developers follow commenting conventions in writing code comments in Java and Python?}

Our initial results show that the majority of style guidelines propose more content-related conventions than other types of conventions, but compared to the different types of content developers actually embed in comments~(\cite{Pasc17a,Zhan18a,Rani21d}), and the concerns they raise on online platforms (\eg StackOverflow or Quora) regarding comment conventions~\cite{Rani21e}, it is clear that existing conventions are neither adequate, nor precise enough.
On the other hand, these style guidelines often include conventions that are not relevant or applicable in many cases, leading developers to ignore them.

When the conventions are applicable, developers often follow the writing style and content conventions (80\% of comments), but violate structure conventions in Java and Python class comments (nearly 30\% of comments), confirming the previous results for Smalltalk by Rani~\etal~\cite{Rani21b}.
Although the project-specific guidelines provide very few additional class comment conventions, these conventions are followed more often compared to the conventions suggested by the standard guidelines both in Java and Python class comments. 
The data related to RQ$_1$ and RQ$_2$ is given in the replication package.\smallfootnote{\url{https://doi.org/10.5281/zenodo.5296443}}

Our results highlight the mismatch between conventions suggested by the style guidelines but not followed by developers in their projects and vice versa.
Verifying this mismatch is currently not well-supported by tools or linters.
The linters currently support very limited comment conventions \cite{Smit11a,Simm20a}, \eg they check the presence or absence of code comments, or their formatting with code but not their content.
Our results indicate the need to conduct extensive studies on (i) which comment-related conventions linters provide, (ii) how well linters cover comment conventions from various style guidelines, and (iii) building tools and techniques to reduce the mismatch between developer commenting practices and style guidelines.

\section{Study Design}
\seclabel{study-design}
\textbf{Data collection.} 
Rani \etal have previously investigated the adherence of Smalltalk class comments to the conventions provided by a default template~\cite{Rani21b}.
To verify their results across other languages, we analyze class comments in Java and Python.
Rani \etal also investigated class comments of diverse Java and Python projects that vary in terms of size, domain, and contributors~\cite{Rani21d}.
We use their dataset of Java and Python class comments for our analysis (RQ1 and RQ2).
\tabref{tab-project-guidelines} shows the list of these projects in the column \emph{Project} for the selected languages shown in the column \emph{Language}.

\paragraph{Comment conventions in Style Guidelines (\textbf{RQ$_1$})}
The goal of this RQ is to investigate the type of comment conventions (or rules) various style guidelines suggest. 
The term \emph{comment convention} refers to suggestions or rules about various aspects of comments, such as syntax, formatting, content, or writing style.
First, we analyze the coding style guidelines of the projects (listed in \tabref{tab-project-guidelines}) which correspond to the Java and Python projects in the Rani \etal dataset~\cite{Rani21d}.
Then we extract the comment-related rules from these guidelines.
Each project might refer to project-specific guidelines in addition to the standard coding guidelines to customize its coding style.
The standard guidelines are used as the baseline guidelines in the majority of the projects and are often provided by the programming language community such as Oracle, PEP, or organizations such as Google, Apache.
In contrast, the project-specific guidelines scope to the project and extend, clarify, or conflict the standard guidelines such as Pandas.\smallfootnote{\url{https://pandas.pydata.org/pandas-docs/stable/development/contributing_docstring.html}} 
\tabref{tab-project-guidelines} shows if a project supports project-specific commenting guidelines (\checkmark) or not (\texttimes) in the column \emph{Project guideline}, in addition to the standard guidelines the project mentions on its web page (listed in the column \emph{Standard guideline}).

\begin{table}[tbh]
    \centering
    \caption{Overview of the selected projects and their style guidelines.}
    \tablabel{tab-project-guidelines}
    \begin{tabular}{llll}
    \hline
    \textbf{Language} & \textbf{Project} & \textbf{Project guideline} & \textbf{Standard guideline} \\ \hline
    Java & Eclipse & \checkmark & Oracle \\
         & Hadoop      & \checkmark & Oracle \\
         & Vaadin      & \checkmark & Oracle \\
         & Spark       & \checkmark  & Oracle \\
         & Guava       & \texttimes  & Google \\
         & Guice       & \texttimes  & Google \\ 
    \hline
    Python  & Django   & \checkmark & PEP8/257 \\
            & Requests & \checkmark & PEP8/257 \\
            & Pipenv   & \texttimes  & PEP8/257 \\
            & Mailpile & \texttimes  & PEP8/257 \\
            & Pandas   & \checkmark & Numpy \\
            & iPython  & \checkmark & PEP8/257, Numpy \\
            & Pytorch  & \checkmark & {Google} \\ \hline
    \end{tabular}
\end{table}

As within a style guideline, comment conventions can be scattered across multiple paragraphs, we scan all sentences and select those that mention any convention about comments.
A rule can target various types of comments, such as class, method, package, or inline comments, or part of comments (\eg summary, parameters).
In case a sentence targets multiple comment types, we split the rule for each type.
In total, we collected 600 comment-related rules.
We organized all rules into a taxonomy of five main categories: \emph{Content}, \emph{Structure}, \emph{Formatting}, \emph{Syntax}, and \emph{Writing Style}.
If a rule does not fit any of these categories, we put it into the \emph{Other} category.
The rationale behind the taxonomy is that the approaches evaluating comment quality can focus on a specific aspect of comments they want to evaluate and improve.
Categories such as \emph{Content} and \emph{Writing Style}, are considered important by Rani~\etal in their work on evaluating Smalltalk comments~\cite{Rani21b}.

According to Rani~\etal, the \emph{Content} category contains the rules that describe which type of information the comment should contain while the \emph{Writing Style} category contains natural-language specific rules, such as grammar, punctuations, and capitalization.
We added three more categories, following their methodology, to cover other aspects of comments.
The \emph{Formatting} category deals with the rules related to indentation, blank lines, or spacing.
It often complements \emph{Structure} category conventions.
The category \emph{Structure} contains the rules about organizing the text, or location of the information in comments. For example, how the tags/sections/information should be ordered in the comments.
The \emph{Syntax} category focuses on the syntax to write a specific type of comments, for instance, which symbol to use to denote comments.
Then, we analyze the frequency of these categories in the style guidelines to answer the RQ$_1$.
\paragraph{Adherence of comments to conventions (\textbf{RQ$_2$})}
The goal of this RQ is to verify whether or not developers follow the comment conventions, \ie the rules identified in the previous RQ, in practice in their projects.
Currently, there are no tools available that automatically check all comment types against all rule types, therefore, from each project we manually validate a sample of class comments against all class comment related rules extracted from its standard and project-specific guidelines shown in \tabref{tab-project-guidelines} (390 rules out of 600 rules).
For the scope of this work, we focus on the rule types that require manual validation due to limited tool support \ie all types of rules other than \emph{Formatting},
thus, we validate 270 rules against the sample class comments.

We use the dataset by Rani~\etal that provides sample class comments of the selected projects shown in \tabref{tab-project-guidelines}~\cite{Rani21d}.
They selected a statistically significant sample of class comments from all class comments of each project with 95\% confidence and 5\% margin of error, resulting in a total of 700 class comments for both languages.
In case a comment follows a particular rule, we label the rule as \emph{followed}, otherwise as \emph{not followed}.
There are often cases where a rule is not applicable to the comment due to the unavailability of that information in the comment, \eg the rules verifying syntax, content, or style of the version information in a class comment cannot be checked if the version information is not mentioned in the comment.
For such cases, we label such rules as \emph{not applicable} to the comment.
We exclude a few rules for now that cannot be verified with the current dataset due to the abstract nature of a rule, the unavailability of the symbols that denotes the class comment, or code associated with the class, \eg to verify the Oracle rule \crule{for the @deprecated tag, suggest what item to use instead}, class comment alone is not enough and require code of the class to verify the replacement item.
We plan to extend the dataset with more comment types and the required data.

We measure how many comments follow a particular rule and how many do not.
One author labels the comments, and the second author reviews the labeled comments.
In cases where they do not agree, the third author is consulted, and conflicts are resolved using the majority voting mechanism (Cohen's kappa=0.80).

\section{Early Results}
\seclabel{results}
\paragraph{Comment conventions in Style Guidelines (\textbf{RQ$_1$})}

\begin{figure}[h!]
    \centering
    \includegraphics[width=\linewidth]{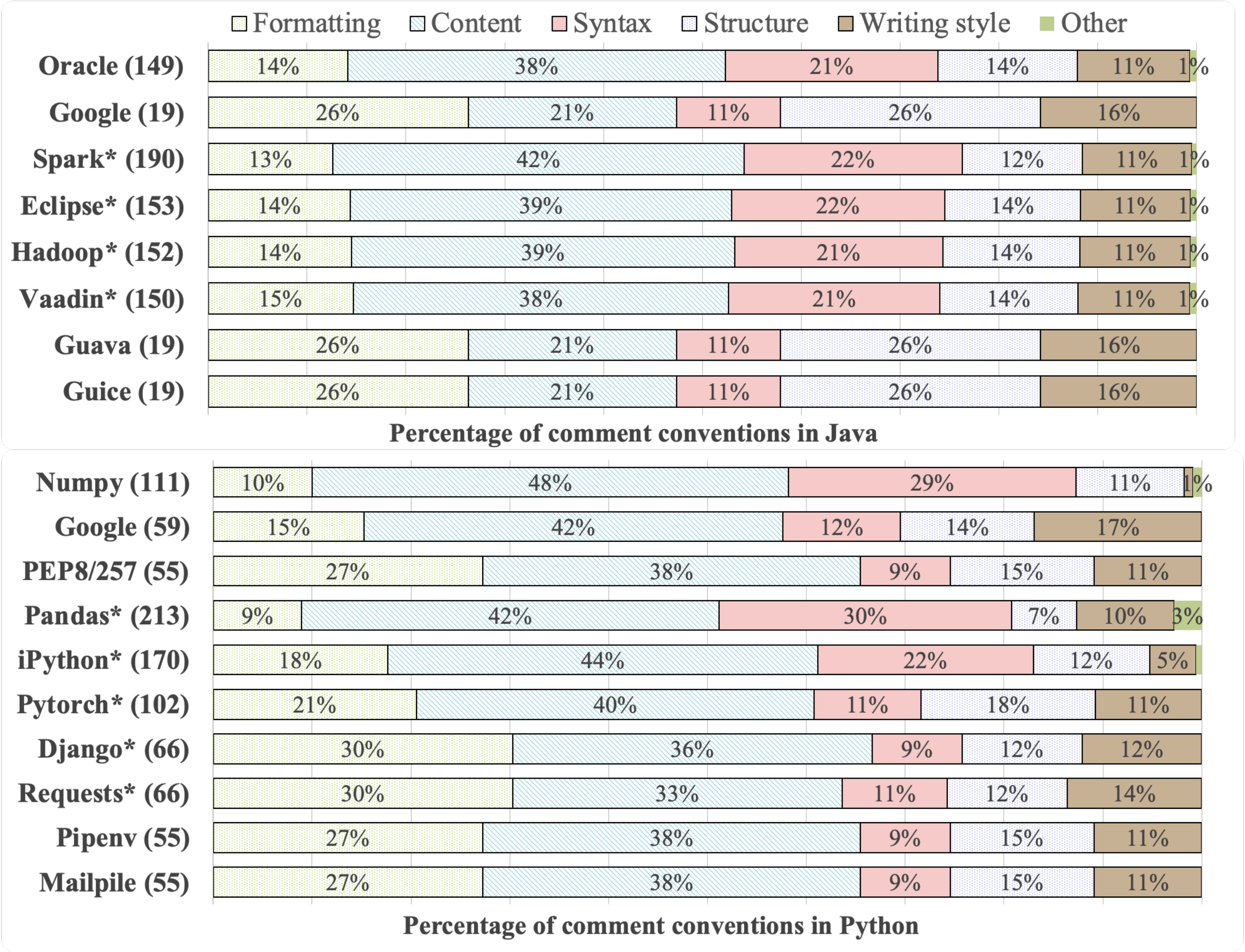}
    \caption{Types of conventions in Java and Python guidelines}
    \figlabel{java_python_convention_distribution}
\end{figure}

\figref{java_python_convention_distribution} shows the total number of conventions for each standard guideline (Oracle and Google) and project-specific guidelines (including conventions from the standard guidelines) on the y-axis.
The x-axis indicates the ratio of conventions belonging to a particular category from our taxonomy.
Our results show that the majority of style guidelines present more rules about the content to write (\emph{Content}) in comments, except for the Google style guideline in Java, which contains more rules on how to format and structure comments (\emph{Formatting}, \emph{Structure}).
Since the Oracle guideline is used as a baseline in several Java projects, project-specific guidelines suggest few additional comment conventions, and these conventions often either conflict with or clarify the standard guidelines.
For example, the conventions, such as \emph{line length limit} and \emph{indentation with two spaces, four spaces, or tab}, are often among such additional and conflicting rules across projects. Identifying such rules and ensuring they are configured properly in tools can help developers in following them automatically.

\figref{java_python_convention_distribution} also shows the distribution of rule types for Python style guidelines.
Numpy and the projects following it (Pandas and iPython) contain the most rules about what type of information to write in comments and how to write it, compared to other standard guidelines, such as those of Google, PEP, and Oracle.
For example, the Numpy guideline suggests writing a short and extended summary of the class, usage examples, notes and warnings in a class comment, and provides syntax and style conventions to write these types of information.
Class comments of iPython and Pandas contain all of these information types and follow the syntax conventions to write them.
Interestingly, developers write such types of information in all other projects~\cite{Zhan18a, Rani21d} regardless of whether the project guideline suggest or not, but they are writing these information types in inconsistent ways.
Previous comment analysis studies for Java and Python also show that developers embed other, different types of information in comments, such as \emph{Usage}, \emph{Expand}, \emph{Rationale}, or \emph{Pointer}, but we do not find conventions in the corresponding style guidelines (Google, PEP, Oracle) to write such types of information~\cite{Pasc17a,Zhan18a}.

We observe that even though style guidelines are intended to encourage and help developers to write good comments for all code entities, comment conventions are scattered across multiple sources, documents, and paragraphs.
Thus, it is not always easy to locate conventions particular to one entity (class, function, inline), causing developers to seek conventions using online sources~\cite{Rani21e}.

\noindent\fbox{%
\parbox{\linewidth}{%
\fontsize{9}{10}\selectfont
\textbf{Finding.} 
The majority of the style guidelines propose more content-related conventions than other types of conventions, but they are not easy to locate in the style guidelines, and  do not always match developer commenting practices.
}%
}

\noindent\fbox{%
\parbox{\linewidth}{%
\fontsize{9}{10}\selectfont
\textbf{Finding.} 
The Numpy style guideline provides more rigorous content conventions for comments compared to other style guidelines, such as Oracle, PEP257, or Google.
}%
}\\

\paragraph{Comment conventions in Style Guidelines (\textbf{RQ$_2$})}

\begin{figure}[h!]
    \centering
    \includegraphics[width=\linewidth]{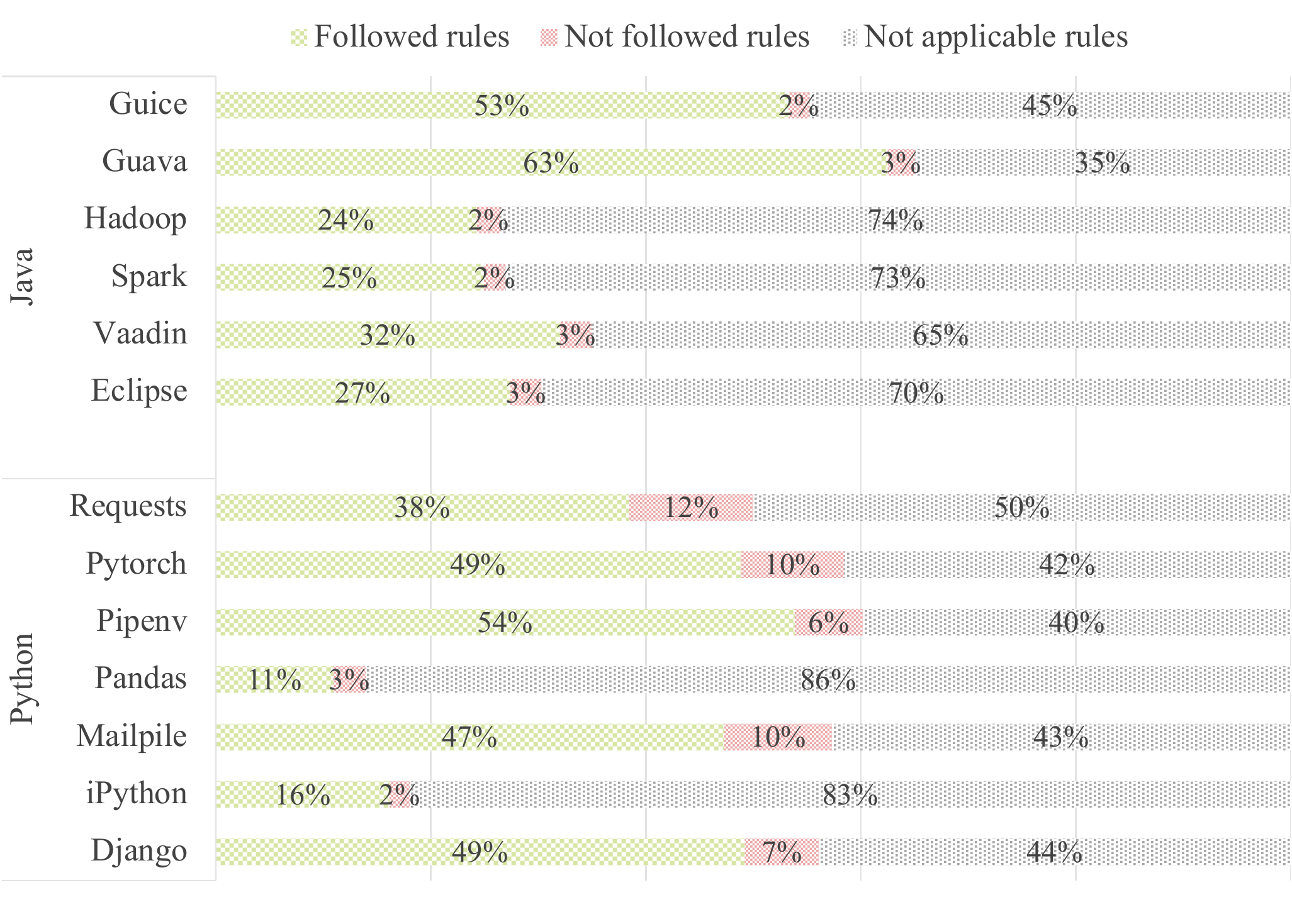}
	\caption{Percentage of comments that follow rules, do not follow them, or to which rules are not applicable.}
    \figlabel{java_python_rules_followed_notFollowed}
\end{figure}

\figref{java_python_rules_followed_notFollowed} shows the distribution of comments within each project that follows rules, do not follow them, or to which the rules are not applicable.
For example, in Eclipse on average 27\% of the comments follow the rules, whereas 3\% of comments violate the rules, and 70\% of the comments do not have enough relevant information within them to check them against a rule.
High ratio of non applicable rules (shown in \figref{java_python_rules_followed_notFollowed}) to selected comments 
indicates that the style guidelines suggest various comment conventions, but developers rarely adopt them while writing comments.
For instance, the Oracle rules in Java, such as \crule{use FIXME to flag something that is bogus or broken}, or \crule{use @serial tag in class comment} are
not applicable on comments due to unavailability of FIXME or @serial information in comments and thus 
showing the developers interest in not adopting such rules.
Similarly, some rules in Python, such as \crule{Docstrings for a class should list public methods and instance variables} are also rarely adopted.

\noindent\fbox{%
\parbox{\linewidth}{%
\fontsize{9}{10}\selectfont
\textbf{Finding.} 
Style guidelines suggest various comment conventions, but developers do not or rarely adopt them while writing comments.
}%
}

 \begin{figure}[tbh]
    \centering
     \includegraphics[width=\linewidth]{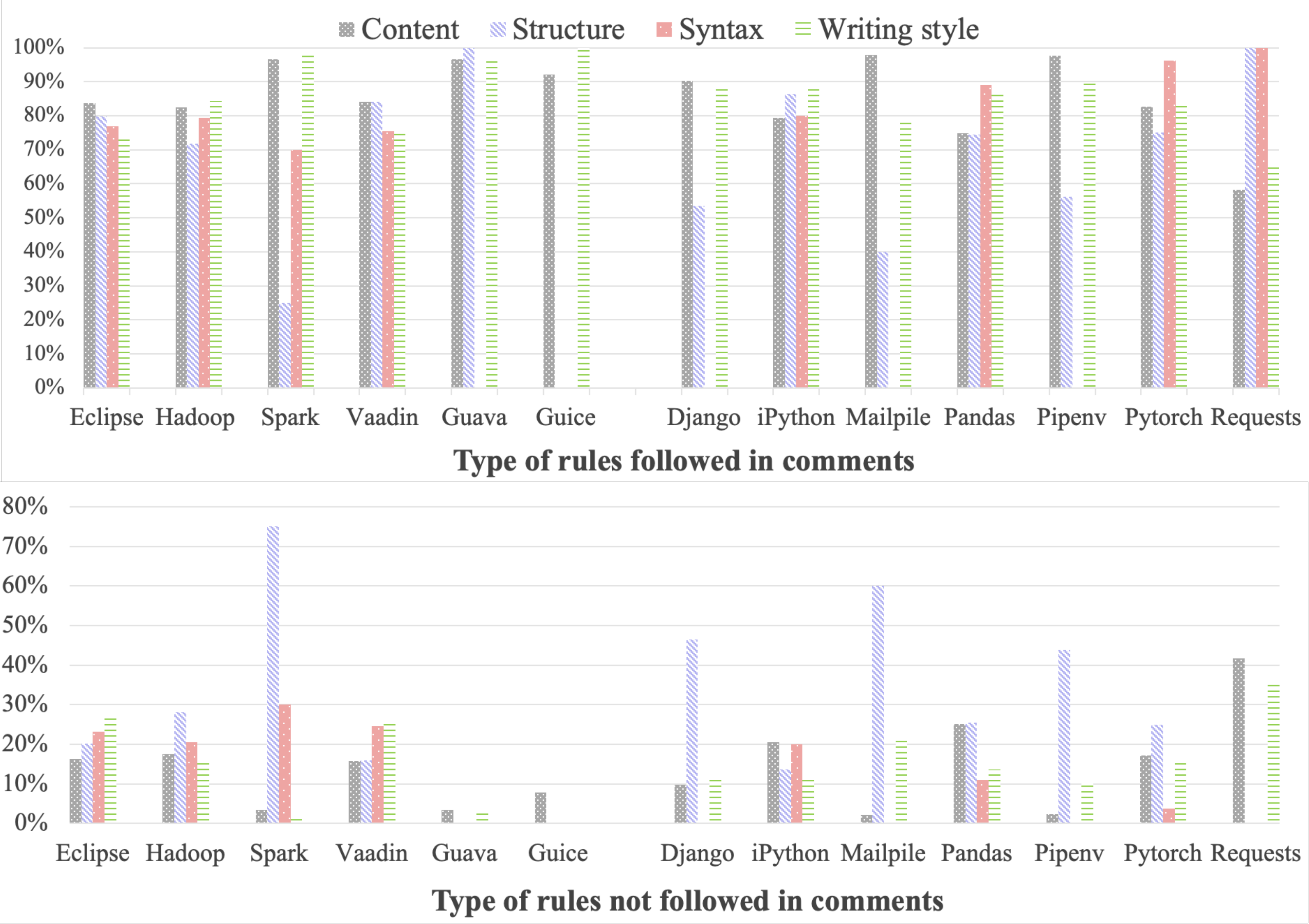}
     \caption{Types of rules followed or not in Java and Python projects}
     \figlabel{java_python_validated_comments_distribution}
 \end{figure} 

Of the rules that are applicable as shown in \figref{java_python_validated_comments_distribution}, writing style and content rules were more often followed than syntax and structure rules, confirming previous results for Smalltalk~\cite{Rani21b}. It shows that developers are interested in writing informative and consistent comments.

\noindent\fbox{%
\parbox{\linewidth}{%
\fontsize{9}{10}\selectfont
\textbf{Finding.} 
Compared to Python, Java class comments violate rules less often (as shown in \figref{java_python_validated_comments_distribution}).
}%
}

\noindent\fbox{%
\parbox{\linewidth}{%
\fontsize{9}{10}\selectfont
\textbf{Finding.} 
Class comments in Java and Python often follow writing style and content conventions (80\% of comments), but violate structure conventions (30\% of comments).
}%
}\\

As discussed before, some rules are often followed, while there are others that are frequently violated.
For instance, the syntax rule \crule{separate the paragraphs with a \texttt{<p>} paragraph tag} in Spark is violated often.
Similarly, in Pandas the rule \crule{a few sentences giving an extended summary of the class or method after the short (one-line) summary} is often followed but the rule \crule{there should be a blank line between the short summary and extended summary} is often violated.
Such conventions can be further investigated by surveying developers to know the specific factors, such as the usage of linters for comments, team strictness, or developer awareness behind these explicit instances of rule adherence or violation.
Although the project-specific guidelines, as shown in \tabref{tab-project-guidelines}, provide few additional conventions, these conventions are followed more often in the projects compared to the conventions provided by their standard guidelines.
Specifically, 85\% of Python class comments and 89\% of Java class comments follow the project-specific conventions, whereas 81\% of their comments follow the conventions from the standard guidelines.
One such example is, the rule \crule{Do not use @author tags} is specific to Hadoop and in contrast with the Oracle style guideline, but it is always followed in Hadoop comments. It would be an interesting future work to explore the reasons behind such conventions.

\vspace{2mm}
\noindent\fbox{%
\parbox{\linewidth}{%
\fontsize{9}{10}\selectfont
\textbf{Finding.} 
Project-specific class comment conventions are followed more often than the conventions suggested by the standard guidelines.
}%
}\\

\section{Implication \& Related Work}
\label{sec:related-work}

\textbf{Impact of commenting conventions}. Coding style guidelines impact program comprehension and maintenance activities.
However, not all conventions from the guidelines have the same impact on these activities.
Smit \etal \cite{Smit11a} ranked 71 code conventions that are most important to maintainable code.
However, they accounted only for missing or incomplete Javadoc comments on public types and methods, and did not account for other comment-related conventions, especially about their content.
Similarly, most previous work has focused on building tools for formatting and naming conventions for code entities, while being very limited on comment conventions~\cite{Alla14a,Arai14a}.
We provide a dataset of 700 labeled class comments and 600 comment conventions (taxonomy) for Java and Python.
This dataset can help researchers rank the specific comment conventions to find out their importance, and impact on the program comprehension and maintenance activities, and thus help in developing comment quality tools based on the supervised machine-learning approaches.\\
\textbf{Comment generation}. To reduce developer effort in writing comments, various researchers have proposed to generate comments automatically.
Moreno \etal proposed a template-based approach to generate class comments in Java~\cite{More13c}.
Given the importance of including developer commenting practices in such approaches, and the impact of a template on developers~\cite{Rani21b}, our results can help researchers design such templates more carefully.\\
\textbf{Adherence of comment conventions}. 
Previous works, including Bafatakis \etal and Simmons \etal, evaluated the compliance of Python code to Python style guidelines~\cite{Bafa19a,Simm20a}. 
However, they included only few comment conventions and missed many other content and writing style conventions.
In our study, we find various comment conventions, such as grammar rules, the syntax of writing different types of information that developers often follow, but which are not covered in such studies.
Rani \etal measured the adherence of Smalltalk class comments to the default comment template and found that developers follow the writing conventions of the template~\cite{Rani21b}.
Java and Python do not provide any default template to write comments but support multiple style guidelines for each project, thus collecting and verifying their comment conventions against comments is more tricky.
We study diverse projects in Java and Python and found that developers follow writing and content conventions more than other types, thus confirming the results of Rani \etal for Smalltalk.

\section{Conclusions}
\label{sec:conclusion}
Given the importance of code comments and consistency concerns in projects, we study various style guidelines and diverse open-source projects in the context of comment conventions.
We highlight the mismatch between what conventions the style guidelines suggest for class comments, and how often developers adopt and follow them, and what conventions developers follow in their class comments but which are not suggested or mentioned by the style guidelines. However, identifying automatically this mismatch is not yet fully achieved. This indicates the need to further automate the software documentation field.
Our results also indicate the need to conduct extensive studies on various linters or quality assessment tools to know the extent they cover various comment conventions and improve them for the missing conventions.
In this direction, we provide a dataset of labeled conventions against comments, and the methodology to extract conventions from the guidelines and verify them against comments. This can help in expanding the scope of the work.
We plan to expand the study to other types of comments, conventions, and programming languages to generalize the current results and design the tools accordingly.
Having the tools to automatically assess the documentation quality at each development stage can help developers in maintaining overall high-quality documentation.

\section{Acknowledgement}
{We gratefully acknowledge the financial support of the Swiss National Science Foundation for the project
``Agile Software Assistance'' (SNSF project No.\ 200020-181973, Feb 1, 2019 - Apr 30, 2022). Bergel is also grateful to Lam Research and the ANID FONDECYT Regular 1200067 for partially sponsoring the work presented in this paper.} 

\bibliographystyle{IEEEtran}
\bibliography{scg,references}
\end{document}